# Diagonal and off-diagonal thermal conduction with resonant phonon scattering in Ni$_3$TeO$_6$

Heejun Yang[1], Xianghan Xu[2,3], Jun Han Lee[4], Yoon Seok Oh[4], Sang-Wook Cheong[2,3], and Je-Geun Park[1,5,$]

[1]Department of Physics and Astronomy, Seoul National University, Seoul 08826, Republic of Korea

[2]Rutgers Center for Emergent Materials, Rutgers University, Piscataway, NJ, USA

[3]Department of Physics & Astronomy, Rutgers University, Piscataway, NJ, USA

[4]Department of Physics, Ulsan National Institute of Science and Technology, Ulsan 44919, Republic of Korea

[5]Institute of Applied Physics, Seoul National University, Seoul 08826, Republic of Korea

[$]jgpark10@snu.ac.kr

## Abstract

The coupling between phonon and magnon is ubiquitous in magnetic materials and plays a crucial role in many aspects of magnetic properties, most notably in spintronics. Yet, this academically and technologically interesting problem still poses a severe challenge to a general understanding of the issue in certain materials. We report that Ni$_3$TeO$_6$ exhibits clear evidence of significant magnon-phonon coupling in both longitudinal thermal conductivity ($\kappa_{xx}$) and thermal Hall coefficient ($\kappa_{xy}$). The Debye-Callaway model, a phenomenological description for phonon heat conduction, can explain the measured magnetic field dependence of $\kappa_{xx}(H)$: phonon scattering from spin fluctuation in the paramagnetic phase and additional scattering due to magnon-phonon coupling in the collinear antiferromagnetic phase. We further suggest that a similar approach could be applied to understand the finite $\kappa_{xy}$ values in Ni$_3$TeO$_6$.

# I. Introduction

Phonons are the dominant heat carrier and usually determine the overall thermal transport properties of most materials. The Debye-Callaway model [1], a standard phenomenological model, captures the essential temperature dependence. But there has been a long-standing question about how these phonons can interact with magnons, another fundamental quasiparticle in magnetic materials—this question goes back to the seminal paper by Kittel [2]. Despite the natural appeal of the idea, one has to wait for a considerable time before seeing actual experiments done with access to detailed precise measurements of both phonons and magnons [3,4].

These recent works primarily concern the case in which the bands of magnon and phonon get coupled to each other. But there can be another case of the magnon-phonon coupling, which is the resonant scattering of phonon via magnon [5–7]. The latter case is relatively rare because it requires a more stringent condition that magnons should have a large density of states at the right energy for the coupling to realise. Recently, the magnon-phonon coupling has renewed interest since the latest theoretical studies suggest that the magnon-phonon coupled term can produce nontrivial Berry curvature to its original Hamiltonian in several magnetic materials [8–11].

Observing Berry phase effects has been regarded as one piece of evidence for verifying the existence of novel quasiparticle [12,13], magnon topology [14–16], and topological phase transition [17] in magnetic insulators. For this reason, the thermal Hall effect, the thermal analog of the electric Hall effect, was proposed as one experimental method to detect such topological effects of the spin system directly. In principle, phonon can also possess Berry curvature with possible phonon Hall effect (PHE) [18,19], which implies a thermal Hall effect mainly governed by phonon. However, the typical size of PHE was expected to be negligible compared to its magnetic origin, as in $Tb_3Ga_5O_{12}$ [20], the first material showing PHE. Indeed, several successful thermal Hall effect measurements from magnon [21–24], nontrivial spin excitation in frustrated lattice [25–28], and Majorana fermion [29–31] were reported so far without considering the phonon contribution.

However, there has been a new twist in these thermal Hall effect studies with a series of unexpectedly significant PHE: multiferroicity [32], quantum paraelectricity [33], structural domain [34], and pseudogap phase in cuprate [35–37]. These latest results all seem to point towards possibly genuine phonon effects. One common feature of these systems is that the overall temperature dependence between longitudinal thermal conductivity ($\kappa_{xx}$) and thermal Hall coefficient ($\kappa_{xy}$) is quite similar to one another. With an apparently dominant phonon contribution to $\kappa_{xx}$ for insulating system, the similar temperature dependence between $\kappa_{xx}(T)$ and $\kappa_{xy}(T)$ can be interpreted as that $\kappa_{xx}(T)$ and $\kappa_{xy}(T)$ would share the same origin, a likely candidate being phonons [34–38]. Moreover, recent studies showed that such similar temperature dependence between $\kappa_{xx}(T)$ and $\kappa_{xy}(T)$ can also be found in other more exotic systems: e.g., the Kitaev quantum spin liquid candidate α-RuCl$_3$ [39] and kagome antiferromagnet Cd-kapellasite [40]. They suggested that sizable $\kappa_{xy}$ values found in the experiments might not only be solely from Majorana fermion or other magnetic origin, but also from phonon [39–41]. Unfortunately, estimating the phonon contribution in $\kappa_{xy}$ is still hard with detailed mechanisms for PHE wanting [18,19,42–44].

Ni$_3$TeO$_6$ (NTO) can be one good example of the PHE since the specific heat of NTO remains nearly constant with various magnetic fields applied along the crystallographic *c*-axis [45], implying phonon dominant $\kappa_{xx}$ in NTO [32]. It is also noteworthy that NTO is a polar magnet with other interesting properties: huge magnetoelectric effect [46] and clear phonon peak shift in infrared spectroscopy across the antiferromagnetic (AFM) phase transition [47], indicating significant spin-phonon coupling. All these suggest a possible PHE in NTO. There are also two distinct magnetic structures below the Neel temperature ($T_N$): a collinear AFM structure with an easy axis along the *c*-axis [48] in the absence of the magnetic field (Fig. 1(a)) and an incommensurate conical spiral structure [49] of spin-flopped phase above the critical field ($\mu_0 H_c$ ~ 8.5 T) applied along the *c*-axis [46] (Fig. 1(b)). We anticipate that this magnetic phase transition could affect both $\kappa_{xx}$ and $\kappa_{xy}$, giving another clue for PHE.

In this paper, we report both in-plane $\kappa_{xx}$ and $\kappa_{xy}$ of NTO with the magnetic field applied along the *c*-axis up to 14 T. $\kappa_{xx}(T)$ seems to follow the conventional behavior of phonon thermal conductivity with pronounced suppression around $T_N$ due to paramagnetic spin fluctuations. Observed $\kappa_{xy}(T)$ is still finite up to high temperature, twice of $T_N$, indicating that the PHE scenario is more appropriate than magnon. Moreover, the overall temperature dependence of $\kappa_{xy}(T)$ is similar to $\kappa_{xx}(T)$, further suggesting the PHE in NTO. On the other hand, $\kappa_{xx}(H)$ shows two opposite-field dependences between the paramagnetic and AFM phases: slight increasing $\kappa_{xx}(H)$ in $T>T_N$, and more complex decreasing $\kappa_{xx}(H)$ in $T<T_N$. We used the Debye-Callaway model including the resonant phonon scattering process [1,50–53] from magnon to find that it gives a fair phenomenological understanding of $\kappa_{xx}$ for $T<T_N$ and $H<H_c$. We propose that such magnon-phonon interaction would also be a dominant factor for $\kappa_{xy}$.

## II. Experimental Methods

Ni$_3$TeO$_6$ single crystals were grown by a flux method modified from a previous report [54]. Stoichiometric Na$_2$CO$_3$ and TeO$_2$ powders were mixed and sintered at 850 ℃ for 10 hours to make a Ni$_3$TeO$_6$ polycrystalline precursor. For single crystal growth, powders of Ni$_3$TeO$_6$ : V$_2$O$_5$ : TeO$_2$ : NaCl : KCl = 1 : 1.5 : 3 : 3 : 1.5 in the molar ratio were mixed and filled into a platinum crucible. The crucible was kept at 850 ℃ for 10 hours before being slowly cooled to 500 ℃ at a 2 ℃/h rate, after which the heaters were switched off for natural cooling to room temperature. Ni$_3$TeO$_6$ crystals with a typical size of 1.5 mm * 1.5 mm * 0.1 mm were mechanically separated from the product after overnight bathing in hot 1M NaOH. The chirality of Ni$_3$TeO$_6$ crystals was determined by a polarized-light optical microscope [55]. Fig. 1(c) and (d) show that the Ni$_3$TeO$_6$ sample used in this study consists of a single chiral domain.

DC magnetic susceptibility ($\chi_{dc}$) was measured by Quantum Design Magnetic Property Measurement System. Thermal conductivity was measured using a homemade setup, which works based on a conventional steady-state method with one heater and three thermometers. To minimize errors from strong magnetic fields and self-heating of thermometers, homemade SrTiO$_3$ capacitors were adopted as thermometers with careful in-situ calibration [56]. As shown in Fig. 1(e), heat current and the magnetic field are applied along the *ab*-plane and the *c*-axis of the sample, respectively. Three thermometers measure temperature differences along *x* ($\Delta T_x = T_1 - T_2$) and *y* ($\Delta T_y = T_3 - T_2$) directions, which can be converted into $\kappa_{xx}$ and $\kappa_{xy}$, respectively [57].

During the sample preparation, contact misalignment between two transverse contacts on sample ($T_2$ and $T_3$) is inevitable. Thus, $\Delta T_y$ can contain both true Hall response ($\delta T_y$) and small amount ($|\alpha| \ll 1$) of longitudinal component ($\alpha \Delta T_x$), i.e. $\Delta T_y = \delta T_y + \alpha \Delta T_x$. Here we should note that $\alpha \Delta T_x$ usually dominates measured $\Delta T_y$ since the typical size of $\delta T_y$ is about 1000 times smaller than $\Delta T_x$ in insulator. To extract the accurate value of $\delta T_y$ from $\Delta T_y$, we applied an antisymmetrization procedure of $\Delta T_y$ with opposite magnetic field directions using the following relation $\delta T_y(+H) = \frac{\Delta T_y(+H) - \Delta T_y(-H)}{2}$, since $\Delta T_x$ is symmetric ($\Delta T_x(+H) = \Delta T_x(-H)$) and $\delta T_y$ is antisymmetric ($\delta T_y(+H) = -\delta T_y(-H)$) to the magnetic field.

## III. Data and Analysis

Fig. 2(a) shows the $\chi_{dc}$ measured with magnetic field of 0.2 T parallel to the *c*-axis after zero-field cooling. With decreasing temperature from 300 K, $\chi_{dc}$ follows the Curie-Weiss law exhibiting a single peak near 52 K and converges toward zero rapidly as the temperature goes 0 K. We determined $T_N$ of NTO as 52 K from the sharp peak in $d\chi_{dc}/dT$ (inset of Fig. 2(a)), consistent with the previous report [46]. The black straight line in Fig. 2(a) is obtained from the Curie-Weiss law with fitting range between 150 and 300 K. The experimental $1/\chi_{dc}$ starts to deviate from the fitting below 100 K, indicating short-range correlations for $T_N < T < 100$ K.

Fig. 2(b) displays $\kappa_{xx}(\mu_0 H = 0, 14$ T$)$ and $\kappa_{xy}(\mu_0 H = 14$ T$)$ as a function of temperature. For $T < T_N$, $\kappa_{xx}(T)$ follows the typical phonon thermal conductivity with a peak around 30 K [58]. However, $\kappa_{xx}(T)$ starts to increase gradually just above the $T_N$, which does not match with the decreasing nature of typical phonon conductivity at higher temperatures. Interestingly, $\kappa_{xx}(T)$ is restored back to the decreasing behavior for $T>130$K, normally expected for phonons. This broad increase of $\kappa_{xx}(T)$ seen for $T_N < T < 130$ K is difficult to understand especially as there exists strong spin fluctuation in the paramagnetic phase, as found in $\chi_{dc}$. Previous studies found that spin fluctuations would be expected to suppress, not boost, $\kappa_{xx}(T)$ since spin fluctuations can be regarded as additional scattering source for phonon heat conduction [59–64]. On the other hand, a magnetic field produces negligible effects to $\kappa_{xx}(T)$ for T>$T_N$ whereas there is overall suppression in $\kappa_{xx}(T)$ for T<$T_N$.

As regards thermal Hall measurement, we found $\kappa_{xy}(T)$ to behave negatively linear with magnetic fields overall temperature range up to twice of $T_N$. For T>$T_N$, finite $\kappa_{xy}(T)$ was observed without further changes even at 120 K. It is well known that magnon Hall conductivity disappears rapidly above a magnetic ordering temperature [21–24]. Therefore, we can assume that $\kappa_{xy}(T)$ in NTO should originate from other sources for T>$T_N$: for which phonon is a natural candidate. Upon cooling from $T_N$, the magnitude of $\kappa_{xy}(T)$ increases rapidly with a peak around 30 K before being suppressed at lower temperature. Hence, both $\kappa_{xx}(T)$ and $\kappa_{xy}(T)$ show a very similar temperature dependence in NTO. This similiarity again reinforces our conclusion that the PHE is most likely to be dominant in the AFM phase of NTO, as in Ref. [34–38].

The detailed field dependences of both $\kappa_{xx}(H)$ and $\kappa_{xy}(H)$ are shown in Fig. 3 (a) and (b), respectively. For T>$T_N$, $\kappa_{xx}(H)/\kappa_{xx}(0)$ increases by less than 1 %, which can be interpreted as phonon scattering due to spin fluctuations in the paramagnetic state: the magnetic field aligns the spin moment, and thereby lowers the phonon scattering rate [23,26,27,40,61,65]. On the other hand, the overall decreasing behavior of $\kappa_{xx}(H)/\kappa_{xx}(0)$ seen for T<$T_N$ indicates that dominant phonon scattering mechanism in the ordered phase is different from those in the paramagnetic phase. As temperature lowers, $\kappa_{xx}(H)$ behaves in a more complex manner, showing an upturn and a sharp suppression in $\kappa_{xx}(H)$ seen around $\mu_0H$~8.5 T, the critical field ($H_c$) of the spin-flop transition [46]. This anomaly can be interpreted as an abrupt change of phonon scattering during the magnetic phase transition, adding further evidence for the strong spin-phonon coupling in NTO. On the other hand, $\kappa_{xy}(H)$ shows linear field dependence for all the measured temperatures (Fig. 3(b)). It is noticeable that at lower temperatures $\kappa_{xy}(H)$ also shows an anomaly at the spin-flop transition, which is similar to $\kappa_{xx}(H)$.

According to the Boltzmann transport equation, phonon thermal conductivity ($\kappa_{xx}^{ph}$) can be expressed in terms of specific heat ($C$), group velocity ($v$), and relaxation time ($\tau$) of phonon: $\kappa \sim \frac{1}{3}Cv^2\tau$. Since the specific heat shows negligible magnetic field effect in NTO [45], we can conclude that $\tau$ should play a more dominant role in the field dependence of $\kappa_{xx}^{ph}$ by assuming negligible field effect in $v$. Hence, we adopted the Debye-Callaway model of Eq. (1), a phenomenological model [1], to analyze the field dependence of $\kappa_{xx}(H)$ in the ordered phase of NTO:

$$\kappa_{xx}^{ph} = \frac{k_B^4}{2\pi^2 v \hbar^3}T^3 \int_0^{\frac{T_D}{T}} \frac{x^4 e^x}{(e^x-1)^2}\tau(\omega, T)dx, \quad (1)$$

where $\tau^{-1}(\omega,T)$ is the scattering rate of phonon, $T_D$ is the Debye temperature, $\omega$ is the frequency of phonon and $x = \frac{\hbar\omega}{k_B T}$ is the phonon energy normalized by the thermal energy. The $v$ is estimated from the Debye model using the following relation: $T_D = v\frac{\hbar}{k_B}(6\pi^2 n)^{1/3}$, where $n$ is the number of atoms per unit volume. $\tau^{-1}(\omega,T)$ can then be approximated by a sum of possible scattering sources following Matthiessen's rule: sample boundary ($\tau_{BD}^{-1}$), linear defects ($\tau_{LD}^{-1}$), point defects ($\tau_{PD}^{-1}$), umklapp process ($\tau_{U}^{-1}$) and resonant phonon scattering process ($\tau_{res}^{-1}$),

$$\tau^{-1}(\omega,T) = \tau_{BD}^{-1} + \tau_{LD}^{-1} + \tau_{PD}^{-1} + \tau_{U}^{-1} + \tau_{res}^{-1}. \qquad (2)$$

Each $\tau_{BD}^{-1}$ [66], $\tau_{LD}^{-1}$ [67], $\tau_{PD}^{-1}$ [68] and $\tau_{U}^{-1}$ [58–60] can be modeled as in Eq. (3),

$$\tau^{-1}(\omega,T) = \frac{v}{d} + A_0\omega + A_1\omega^4 + A_2\omega^2 T \exp\left(-\frac{T_D}{bT}\right) + \tau_{res}^{-1}. \qquad (3)$$

The first four terms in Eq. (3) are enough for obtaining representative phonon thermal conductivity, but these are, a priori, assumed to be field-independent. For field-dependent $\kappa_{xx}$, we thus considered resonant phonon scattering process ($\tau_{res}^{-1}$) as described in Fig. 4(a). In this scenario, the low-lying magnon bands having a higher density of states (DoS) split linearly under the magnetic field with an energy difference of $\hbar\omega_{res}$. We note that our hypothesis of the low-lying magnon band with a large DoS is consistent with the recent spin wave measurement [49]. This process then allows phonons to be scattered by magnons having the specific energy of $\hbar\omega_{res}$ [58]. We chose an empirical formula for $\tau_{res}^{-1}$ of Eq. (4), which has been shown to successfully explain $\kappa_{xx}$ in several complex magnetic insulator [50–53]:

$$\tau_{res}^{-1} = R\frac{\omega^4}{(\omega^2-\omega_{res}^2)^2}\frac{\exp\left(-\frac{\hbar\omega_{res}}{k_B T}\right)}{1+\exp\left(-\frac{\hbar\omega_{res}}{k_B T}\right)}, \qquad (4)$$

where $R$ indicates the strength of resonant phonon scattering.

Linearly splitting magnon bands shown in Fig. 4(b) could be natural for NTO, since the magnon modes ($\mathcal{E}_{\pm}$) in the collinear easy-axis AFM phase can be written as Eq. (5) under the magnetic field applied along the easy-axis [69]. We further note that our scenario of the resonant phonon scattering process from magnon makes sense, given that the decreasing $\kappa_{xx}(H)$ happens only below $T_N$. As a result, we can get $\hbar\omega_{res} = 2g\mu_B\mu_0 H$, where $\mu_B$ is Bohr magneton and $g = 2.26$ is the g-factor of NTO [70].

$$\mathcal{E}_{\pm}(H) = \mathcal{E}(H=0) \pm g\mu_B\mu_0 H. \qquad (5)$$

Although $d$, $b$, $T_D$, $A_0$, $A_1$, $A_2$, and $R$ in Eq. (3) and (4) are in principle free parameters, we can put further constraints on several parameters ($d$, $b$ and $T_D$) for a minimum model. First, we can fix $d$ as 1 mm, the shortest in-plane dimension of the sample: $d$ represents the phonon mean free path due to collisions from the sample boundary. Next, $b$, the characteristic constant of the phonon dispersion, can be fixed using a conventional value of $2\sqrt[3]{N}{\sim}6.21$, where $N$ is the number of atoms in a unit cell [59,60,71]. To determine the appropriate value of $T_D$, we fitted the specific heat data taken from previous studies [48,72]; using the Debye-Einstein model, we estimated $T_D = 470$ K. As a result, the free parameters can be reduced to the only four $A_0$, $A_1$, $A_2$ and $R$: all of which indicate the strength of each scattering process.

The fitting result is shown as black dashed curves in Fig. 3(a), and the best fitting parameters are summarized in Table I. Size of these values are comparable to previous studies, which also employed the Debye-Callaway model in their analysis [50–53,59]. The calculated $\kappa_{xx}^{ph}(H)/\kappa_{xx}^{ph}(0)$ reproduces well the overall behavior of $\kappa_{xx}(H)/\kappa_{xx}(0)$ for $T<T_N$ and $H<H_c$, including the upturn seen for $T<10$ K. It is not surprising to find, though, that the fitting breaks down for $T \geq T_N$ since the magnon will be no longer well-defined in the paramagnetic phase. From this exercise, we can conclude that the resonant phonon scattering model may as well be a reasonable explanation for $\kappa_{xx}(H)$ in NTO. However, we admit that our model cannot explain the data in the higher field region $\kappa_{xx}(H > H_c)$. It is mainly due to the fact that the magnetic structure and Hamiltonian are unknown in the spin-flopped phase yet.

Using the same parameters, we also calculated $\kappa_{xx}^{ph}(T)$ at zero-field shown as a black solid curve in Fig. 2(b). We can see that the measured $\kappa_{xx}(T)$ is suppressed from $\kappa_{xx}^{ph}(T)$ in the temperature range of $T_N<T<130$ K. As previous thermal transport studies pointed out, paramagnetic spin fluctuations could scatter phonons off, resulting in flat-like $\kappa_{xx}(T)$ [59,62–64]. Thus, we can conclude that the significant spin fluctuation and magnon strongly affect the phonon heat transport in NTO.

## IV. Discussion

We now like to discuss the implications of our $\kappa_{xy}$. As there is no good simple model for the PHE yet, we can only make general observations by comparing it with other materials having finite $\kappa_{xy}$. First, the mechanism proposed for the PHE in non-magnetic insulator $SrTiO_3$ is hard to be applied to NTO because it requires a substantial dielectric constant ($\varepsilon \sim 10^4$) with structural domain [34,44] or quantum paraelectricity [33]: none of which can be valid for NTO. Instead, we conjecture that in the paramagnetic phase, the PHE in NTO originates from the secondary effect of significant spin-phonon coupling [20,65]. We also noticed that the size of $\kappa_{xy}(T)$ rapidly increases just below $T_N$ (Fig. 2(b)) and $\kappa_{xy}(H)$ starts to show a hump-like behavior at $H_c$ (Fig. 3(b)); all of which indicates that magnetically ordered phase affects $\kappa_{xy}$ significantly. Since the similar temperature dependence between $\kappa_{xx}(T)$ and $\kappa_{xy}(T)$ implies that both $\kappa_{xx}$ and $\kappa_{xy}$ share the origin and magnon-phonon scattering model describes $\kappa_{xx}$ well, we suggest that the magnon-phonon interaction is also a dominant factor for $\kappa_{xy}$ in the AFM phase.

We also anticipate that the magnon-phonon picture presented here could be used to understand other recent thermal Hall experiments properly. For example, the latest study found that temperature dependences between $\kappa_{xx}(T)$ and $\kappa_{xy}(T)$ in $Cu_3TeO_6$ are very similar each other [38], which indicates PHE. We note that $Cu_3TeO_6$ also shows rapidly increasing $\kappa_{xy}(T)$ for $T<T_N$ [38] similar to NTO. From this, we suggest that magnon-phonon interaction also plays a significant role in $Cu_3TeO_6$ for $\kappa_{xy}(T)$, which is consistent to the inelastic neutron scattering study [73]. Furthermore, we can find that $\kappa_{xx}(H)$ and $\kappa_{xy}(H)$ of NTO are quite similar to those of $Fe_2Mo_3O_8$, another polar collinear AFM along the *c*-axis accompanying the spin-flip transition [32]. We expect that our model can also be used to explain the data in $Fe_2Mo_3O_8$.

To summarize, we measured both $\kappa_{xx}$ and $\kappa_{xy}$ of $Ni_3TeO_6$. We observed finite negative $\kappa_{xy}$ up to two times of $T_N$ and a similar temperature dependence between $\kappa_{xy}$ and $\kappa_{xx}$, which indicates PHE in $Ni_3TeO_6$. The collinear AFM phase has $\kappa_{xx}$ well described by the Debye-Callaway model with resonant phonon scattering from the magnon band. We suggest that the same origin governs both $\kappa_{xx}$ and $\kappa_{xy}$: spin-phonon coupling for $T>T_N$ and magnon-phonon interaction for $T<T_N$. We expect that the PHE from magnon-phonon interaction could be applied to other insulating magnets in the magnetically ordered phase.

**Acknowledgment**

We want to thank Ysun Choi for the helpful discussion. The work at SNU was supported by the Leading Researcher Program of Korea's National Research Foundation (Grant No. 2020R1A3B2079375). The work at Rutgers University was supported by the DOE under Grant No. DOE: DE-FG02-07ER46382. J.H.L. and Y.S.O. acknowledge support from the Basic Science Research Programs through the National Research Foundation of Korea (NRF) (NRF-2020R1A2C1009537).

# Appendix

### 1. SrTiO$_3$ capacitive thermometry for thermal Hall measurement

Both side polished SrTiO$_3$ wafers of 0.1 mm thickness were purchased from Crystal GmbH and cut into size of 1×1 mm$^2$ by diamond wire saw, to make parallel plate geometry. After the 50 nm layer of gold evaporation, silver epoxy was coated on both side. Silver wire of 127 μm diameter was then used to connect sample and thermometers to maximize thermal conductance between them. To minimize heat leak through the thermometers, we adopted phosphor bronze wire of 25 μm diameter as leads, which is known to have poor thermal conductance.

Typical temperature dependence of capacitance for SrTiO$_3$ capacitive thermometer shows monotonic increasing as temperature goes down to 4 K (see Fig. 5(a)). Before thermal Hall measurement, we always performed in-situ calibration under the zero magnetic field at each target temperature to minimize calibration errors further. We did not calibrate our thermometer for different magnetic field values since the dielectric constant of SrTiO$_3$ is known to show negligible magnetic field effect [74].

During the thermal Hall measurement, we applied static magnetic field at each field step and waited around 15 minutes to eliminate possible error from magneto-caloric effect. We also proceeded the control experiment to check uncertainty of our measurement setup by measuring antisymmetrized $\Delta T_y$ without $\Delta T_x$. As shown in Fig. 5(b), the uncertainty is less then 0.1 mK, which is much smaller than typical Hall response with order of 1 mK. Further technical information is available in Ref. [56].

### 2. Fitting procedure for Debye-Callaway model

Before starting the fitting the data, it is better to know how each terms in phonon scattering rate of Eq. (2) manipulate $\kappa_{xx}^{ph}$. At first, $\tau_{BD}^{-1}$ affects $\kappa_{xx}^{ph}$ in low temperature region, since phonon mean free path should be comparable to sample size for significant boundary scattering [58]. Next, presence of $\tau_{LD}^{-1}$ and $\tau_{PD}^{-1}$ give overall suppression of $\kappa_{xx}^{ph}$ for wide temperature range. And $\tau_U^{-1}$ tunes the degree of decreasing behavior of $\kappa_{xx}^{ph}$ for high temperature range. At last, $\tau_{res}^{-1}$ mainly gives field dependence to $\kappa_{xx}^{ph}$ in our model. We assumed all parameters in Eq. (3) and (4) are independent to both temperature and magnetic field.

Considering the above properties of each scattering process, we could roughly determine initial parameter set for $\kappa_{xx}(T, \mu_0 H = 0\text{T})$ without considering $\tau_{res}^{-1}$ first. Then, we turned on $\tau_{res}^{-1}$ to fit both $\kappa_{xx}(T, \mu_0 H = 0\text{T})$ and magneto-thermal conductivity ($\kappa_{xx}(H)/\kappa_{xx}(0)$) simultaneously, by minimizing $\chi^2$. To find best parameter set shown in Table I, we used particle swarm optimization algorithm [75], which is powerful for seeking global mimimum of complex non-linear function with broad parameter space.

# Figures

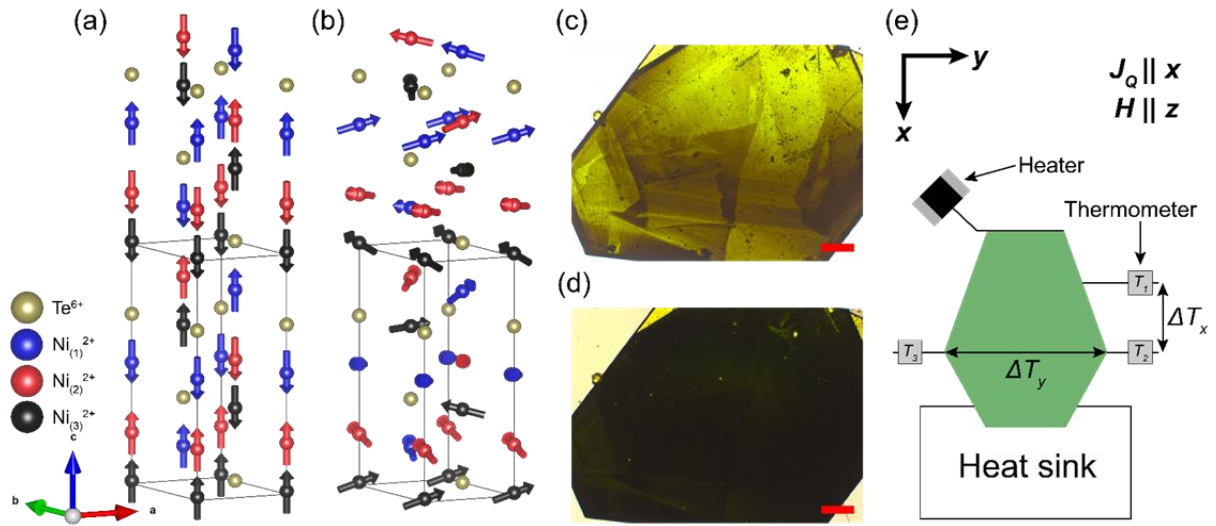

Figure 1 **Magnetic structure and photo of Ni$_3$TeO$_6$ sample and schematic of thermal transport experimental setup** (a) Collinear antiferromagnetic structure of the low-field phase. (b) The incommensurate conical spiral magnetic structure of the spin-flopped phase, as suggested in Ref. [49]. (c) and (d) Images of Ni$_3$TeO$_6$ sample taken using a transmission polarized optical microscope. Scale bars are 200 μm. (c) The sample shows transparent green color with parallel polarizer and analyzer. (d) Slight rotation of the analyzer makes the chiral domain to be visible. The overall dark color indicates a single chiral domain in the sample. (e) Schematic of thermal transport experimental setup. Heat current ($J_Q$) and magnetic field ($H$) are applied along the *ab*-plane and the *c*-axis of the sample, respectively.

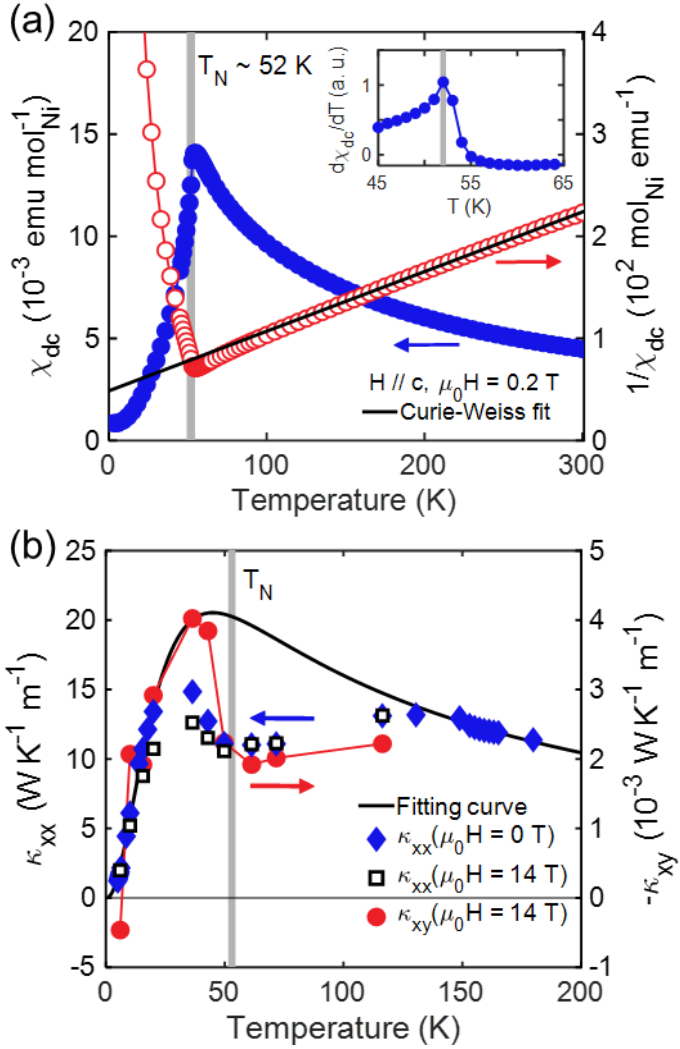

**Figure 2 Measured DC magnetic susceptibility ($\chi_{dc}$) along the *c*-axis, longitudinal thermal conductivity ($\kappa_{xx}$) and thermal Hall coefficient ($\kappa_{xy}$) as a function of temperature.** (a) Blue filled and red open circles display $\chi_{dc}$ and $1/\chi_{dc}$, respectively. The inset shows $d\chi_{dc}/dT$ with a clear transition peak near 52 K, indicating an antiferromagnetic phase transition. A solid black line is obtained from the Curie-Weiss law. (b) Blue diamond and open square show $\kappa_{xx}$ at 0 and 14 T, respectively. Red circle shows $-\kappa_{xy}$ at a magnetic field of 14 T. The solid black curve is obtained from the Debye-Callaway model by fitting the zero-field $\kappa_{xx}$ data.

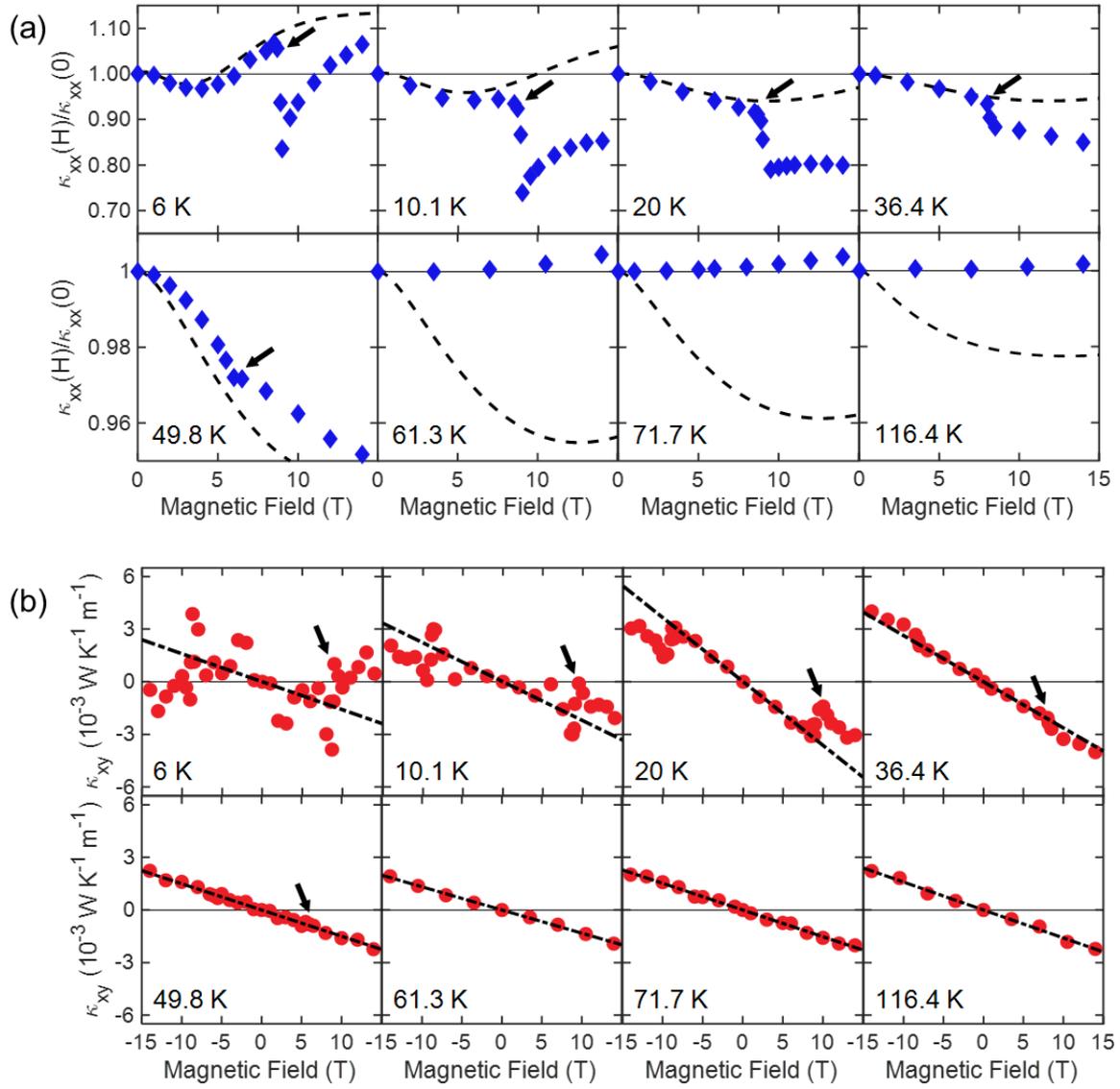

Figure 3 **Measured magneto-thermal conductivity ($\kappa_{xx}(H)/\kappa_{xx}(0)$) and thermal Hall coefficient ($\kappa_{xy}$) as a function of magnetic field.** (a) Blue diamond shows $\kappa_{xx}(H)/\kappa_{xx}(0)$ at various temperature points. Black dashed curves are obtained from the Debye-Callaway model. (b) The red circle shows the field dependence of $\kappa_{xy}$ at various temperatures. Black dash-dotted lines are guides to the eye clarifying the dominant linearity of $\kappa_{xy}$, especially in the low-field collinear antiferromagnetic phase. Black arrows indicate a sharp anomaly on $\kappa_{xx}$ and $\kappa_{xy}$ due to the spin-flop transition.

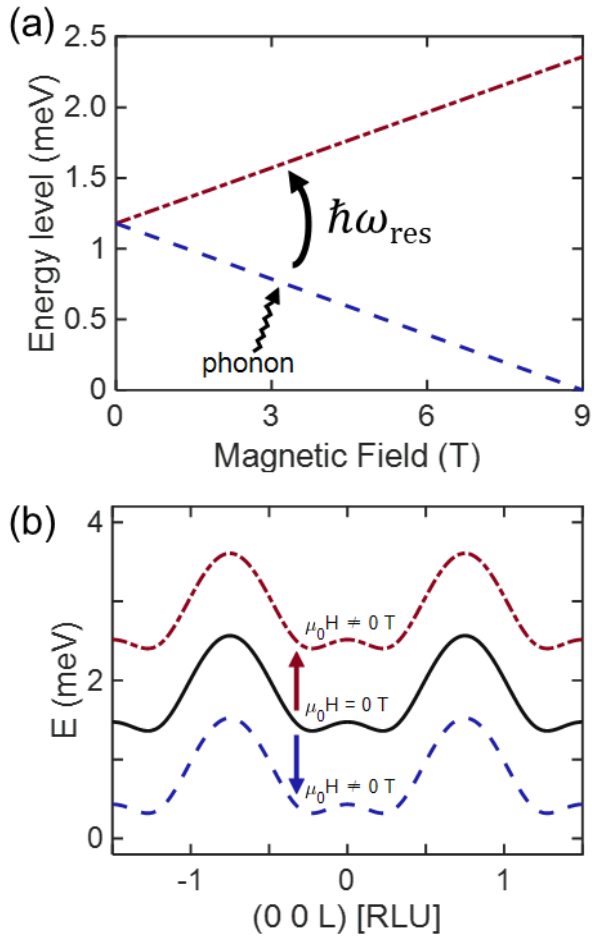

Figure 4 **Schematic of resonant phonon scattering process and effective magnon dispersion of Ni$_3$TeO$_6$ in collinear antiferromagnetic phase.** (a) Two energy levels in low-lying magnon bands get linear split under the magnetic field. $\hbar\omega_{res}$ is the energy difference between lower and upper energy levels of the magnon bands. Blue dashed and red dash-dotted lines represent the lower and upper energy levels, respectively. The following process can describe resonant phonon scattering: a lower energy level absorbs phonon with the energy of $\hbar\omega_{res}$, resulting in excitation of the upper level [58]. (b) Effective magnon dispersion in the low-field phase was obtained from Ref. [49]. The black curve shows the dispersion at a zero magnetic field. With a magnetic field, the magnon dispersion starts to split. The red-dotted and blue dashed curves show the dispersion with a finite field.

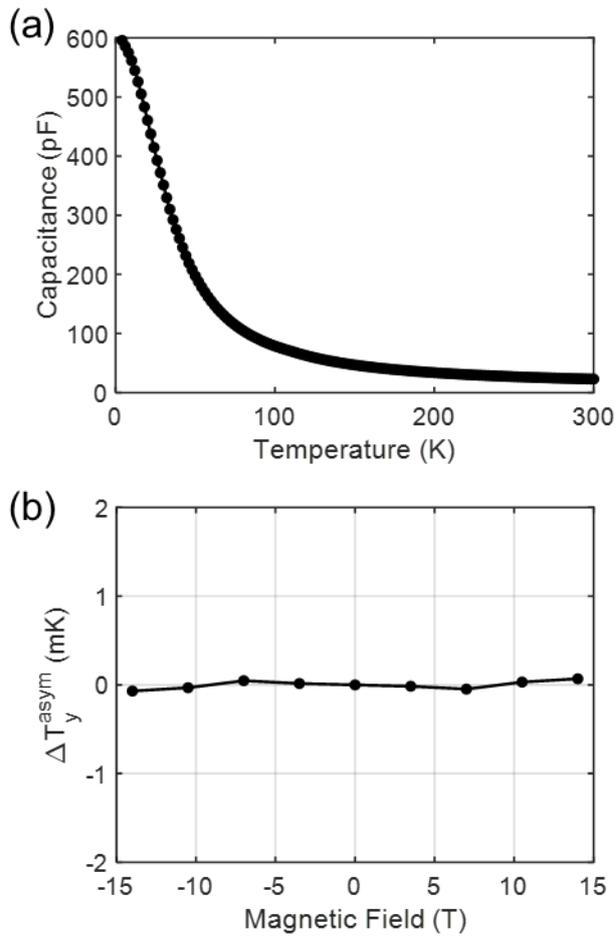

Figure 5 **Typical capacitance value of homemade SrTiO₃ capacitive thermometer and result of control experiment.** (a) Capacitance of SrTiO$_3$ thermometer is given as a function of temperature down to 4 K: the capacitance grows monotonically as the temperature gets lowered. (b) Antisymmetrized transverse temperature difference ($\Delta T_y^{asym}$) obtained from control experiment at 20 K, in which we turned off the heat current along the sample.

| Parameter (unit) | Fitting condition | Value |
|---|---|---|
| $v$ (m/s) | Fixed | 3446 |
| $d$ (mm) | | 1 |
| $T_D$ (K) | | 470 |
| $b$ (unitless) | | 6.21 |
| $A_0$ (unitless) | Free | $3.29 \times 10^{-4}$ |
| $A_1$ (s$^3$) | | $6.44 \times 10^{-43}$ |
| $A_2$ (s·K$^{-1}$) | | $4.04 \times 10^{-18}$ |
| $R$ (s$^{-1}$) | | $1.77 \times 10^{8}$ |

Table I. The best fitting parameters obtained for Debye-Callaway model with their fitting conditions specified.